\documentclass[10pt]{article}
\usepackage{graphicx}
\begin{document}
\centerline{\Large\bf Causality Violation in Non-local QFT \footnote{Talk given at the conference THEP-I held at Indian Institute of Technology, Roorkee, India during 16-20 March 2005}}
\begin{center}
\vskip 0.3in
Satish D. Joglekar\footnote{E-mail: sdj@iitk.ac.in} \\
Department of Physics, IIT Kanpur, Kanpur 208016\\
\end{center}
\vskip 0.3 in
\begin{abstract}
We study the causality violation in the non-local quantum field theory (as formulated by Kleppe and Woodard) containing a finite mass scale
$\Lambda $. We use $\phi ^{4}$ theory as a simple model for study. Starting from the Bogoliubov-Shirkov criterion for causality,
we construct and study combinations of S-matrix elements that signal violation of causality in the one loop approximation. We find that
the causality violation in the exclusive process $\phi +\phi \rightarrow \phi +\phi $
grows with energy, but the growth with energy, (for low to moderate
energies) is suppressed to all orders compared to what one would expect
purely from dimensional considerations. We however find that the causality
violation in other processes such as $\phi +\phi \rightarrow \phi +\phi +\phi +\phi $
grows with energy as expected from dimensional considerations at low
to moderate energies. For high enough energies comparable to the mass
scale $\Lambda $, however, we find a rapid (exponential-like) growth
in the degree of causality violation. We generalize some of the 1-loop  results to all orders. We present interpretations of the results based on possible interpretations of the non-local quantum field theory models.
\end{abstract}

\section{Introduction}

Non-local quantum field theories (NLQFT) have been a subject of wide research since 1950's. The main reason for the interest in early days has been the hope that the non-local quantum field theory can provide a solution to the puzzling aspects of renormalization. The basic idea was that since the divergences in a local quantum filed theory arise from product of fields at an identical space-time point, the divergences of the local quantum field theory would be tamed if the interaction were non-local. In particular, if the interaction scale was typically of the order of $1/\Lambda $, then momenta in loop integrals (Euclidean) would be damped when $|p^2|>>\Lambda ^2$. The early work on NLQFT, starting from that by Pais and Uhlenbeck \cite{PU} and especially that of Efimov and coworkers, has been summarized in \cite{NA}. NLQFT's also have found application towards description of extended particles which incorporates the symmetries of the theory in some (non-local) form \cite{M90}.The non-commutative fields theories, currently being studied \cite{NC}, are a special variant of a NLQFT, as is evident especially in its QFT representation using the star product. In this work, we shall focus our attention on the type of NLQFT's formulated by Kleppe and Woodard \cite{KW92}.

One of the reasons we normally insist on a \emph{local} quantum field theory is because it has microscopic causality, and this ensures causality of the theory.  One of the consequences, therefore, that would be suspected of non-locality would be a causality violation (at least) at the level of the S-matrix. Indeed, since at a given moment, the interaction is spread over a finite region in space, thus covering simultaneously space-like separated points, we expect the interaction could induce non-causality.

In view of the fact that we have not observed large-scale causality violation, it becomes important to generically distinguish between theories exhibiting \emph{classical} violations of causality versus \emph{quantum} violations of causality. As argued in the section~\ref{sec:2}, a violation of causality at the classical level generally has a larger effective range and strength, compared to the quantum violations of causality which are suppressed by ~$g^2/16\pi^2$ per loop. We do not know of large scale causality violations, and as such, it is desirable that the non-local theory has no classical violation of causality. One way known to ensure that there is no classical level of causality violation is to require that the S-matrix of the NLQFT at the tree level coincides with that of the local theory ($\Lambda \rightarrow \infty$).

We shall work in the context of the NLQFT's as formulated by Kleppe and Woodard \cite{KW92}. This form of non-local QFT was evolved out of earlier work of Moffat \cite{M90}, insights into structure of non-local field equations by Eliezer and Woodard \cite{EW} and application to QED by Evens et al \cite{E91}. This formulation has a distinct advantage over earlier attempts in several ways:
\begin{small}  \begin{enumerate}
    \item There are no additional classical solutions to the non-local field equations compared to the local ones. The nonlocal theory is truly a deformation of the local theory and the meaning of quantization, as  a perturbation about the classical, is not altered. This property is not shared by non-commutative field theories.
    \item It has the same S-matrix at the tree level, and thus;
    \item There is no classical violation of causality.
    \item The theory, unlike a higher derivative theory, has no ghosts and is unitary at a \emph{finite $\Lambda$}.
    \item The theory can embody non-localized versions of local symmetries having an equivalent set of consequences.
\end{enumerate}
\end{small}
There are many other reasons for taking interest in these NLQFT's. We have found such a non-local formulation \emph{with a finite $\Lambda $,} very useful in understanding the renormalization program in the renormalizable field theories \cite{j01}. We have shown that this formulation enables one to construct a mathematically consistent framework in which the renormalization program can be understood in a natural manner. The framework does not require any violations of mathematical rigor usually associated with the renormalization program. This framework, moreover, made it possible to theoretically estimate the mass scale $\Lambda $. The nonlocal formulations can
also be understood \cite{j01_2} as an effective field theory formulation of a physical
theory that is valid up to mass scale $\sim $$\Lambda $. In such
a case, the unknown physics at energy scales higher than $\Lambda $ {[}such
as a structure in terms of finer constituents, additional particles,
forces, supersymmetry etc {]} can \emph{effectively be represented} in a \emph{consistent}
way (a unitary, gauge-invariant, finite (or renormalizable) theory)
by the non-local theory. In other words, the nonlocal standard model can serve as such an
effective field theory \cite{j01_2} and will afford a model-independent
way of consistently reparametrizing the effects beyond standard model. It can be looked upon in a number of other ways. One could think the non-locality as representing a form factor with a momentum cut-off $\Lambda $\cite{M90}. One could also think of this theory as embodying a granularity of space-time
of the scale $1/\Lambda $ or as an intrinsic mass scale $\Lambda $
\cite{KW92,js,j01}.\\
A possible "limitation" of the theory is that the theory necessarily has quantum violations of causality \cite{KW92,CO92}; though it can be interpreted as a prediction of the theory. In this work we explore this question with the help of the first calculations for the simplest field theory: the nonlocal version of the $\lambda\phi^4$ theory \cite{JJ}.

In order to study causality violation (CV) in the theory, it is first necessary to formulate quantities that signal CV. We would like to construct quantities that can feasibly be \emph{measured} experimentally. From this view-point, it is appropriate to construct quantities in terms of the S-operator. Bogoliubov and Shirkov \cite{BS} have formulated a \emph{necessary} condition for causality to be preserved in particle physics by the S-operator. This formulation is simple and at the same time extremely general in that,
\begin{small}
\begin{enumerate}
    \item It does not rely on a Lagrangian QFT: it uses only the phenomenologically accessible S-operator.
    \item It uses the most basic notion of causality in a relativistic formulation: A cause at $x$ shall not affect physics at any point $y$ unless $y$ is in the forward light-cone with respect to $x$.
    \item It assumes covariance of observable physics.
    \item It assumes that the S-operator remains unitary in the face of a variable coupling constant\footnote{This requirement, in a theory such as $\lambda\phi^4$ theory, is evidently fulfilled even when $\lambda$ is space-time dependent: the Hamiltonian continues to be Hermitian. In the case of Non-abelian gauge theories, it is possible to construct a BRS-invariant action \emph{with} a \emph{variable} coupling constant. It is best constructed  directly in terms of $\tilde{A}^{\alpha}_{\mu}(x)=g(x){A}^{\alpha}_{\mu}(x)$}.
\end{enumerate}
\end{small}
 The condition is formulated as,\begin{equation}
\begin{array}{ccc}
 \frac{\delta }{\delta g(x)}\left(\frac{\delta S[g]}{\delta g(y)}S^{\dagger }[g]\right)=0 & for & x<\sim y\end{array}
\label{eq:CC}\end{equation}
 where $x<\sim y$ means that either $x^{0}<y^{0}$ or $x$ and $y$ are space-like separated. [In either case, there exists a frame in which $x^{0}<y^{0}$].
 Section~\ref{sec:2} gives a brief qualitative understanding of this relation and how amplitudes indicating causality violation are constructed using this relation.

 In section~\ref{sec:2}, we shall also summarize the essentials of construction of a non-local QFT given a local one. We shall qualitatively discuss causality violation at the classical and the quantum level. In the section 3, we shall give the results for the exclusive processes $ \phi+\phi \rightarrow \phi+\phi$ and $ \phi+\phi \rightarrow \phi+\phi+\phi+\phi$ in the one loop order and also some all-order generalizations of the properties of the former. In the last section, we shall interpret the results. For brevity, we shall leave out many technical details, (for which we shall refer the reader to \cite{JJ}) and prefer to give qualitative elucidations in this article.

\section{Preliminary\label{sec:2}}
In this section, we shall qualitatively discuss the construction of  non-local field theories and the Bogoliubov-Shirkov criterion of causality and make remarks on CV at the classical/quantum level.

\subsection{Non-Local Quantum Field Theory}
We shall present
the construction of the NLQFT as
presented in \cite{KW92}. We start with the local action for a field theory, in terms of a generic field $\phi $, as the sum of the quadratic and the interaction part:
\[
S[\phi ]=F[\phi ]+I[\phi ]\]
and express the quadratic piece as \[
F[\phi ]=\int d^{4}x\phi _{i}(x)\Im _{ij}\phi _{j}(x)\]
We define the regularized action in terms of the smeared field $\widehat{\phi }$,
defined in terms of\footnote{The choice of the smearing operator is the only freedom in the above construction.  For a set of restrictions to be fulfilled by $\mathcal{E}$, see e.g. \cite{CO92}} the kinetic energy operator $\Im _{ij}$ as,\[
\widehat{\phi }\equiv \mathcal{E}^{-1}\phi \, \, \, \, \, \, \, \, \, \, \, \mathcal{E}\equiv \exp [\Im /\Lambda ^{2}]\]
The nonlocally regularized action is constructed by first introducing
an auxiliary action $S[\phi ,\psi ]$. It is given by \[
S[\phi ,\psi ]=F[\hat{\phi }]-A[\psi ]+I[\phi +\psi ]\]
where $\psi $ is called a {}``shadow field'' with an action\[
A[\psi ]=\int d^{4}x\psi _{i}O_{ij}^{-1}\psi _{j};\;\;O\equiv \frac{\mathcal{E}^{2}-1}{\Im } \]
The action of the non-local theory is defined as $
\hat{S}[\phi ]=S[\phi ,\psi ]\Vert _{_{\psi =\psi [\phi ]}}$
 where $\psi [\phi ]$ is the solution of the classical equation $\frac{\delta S}{\delta \psi }=0$

The vertices are unchanged but every leg can connect either
to a smeared propagator\[
\frac{i\mathcal{E}^{2}}{\Im +i\epsilon }=-i\int _{1}^{\infty }\frac{d\tau }{\Lambda ^{2}}exp\{\frac{\Im \tau}{ \Lambda ^{2}}\}\]
or to a shadow propagator {[}shown by a line crossed by a bar{]}

$$\frac{i[1-\mathcal{E}^{2}]}{\Im +i\epsilon }=-iO =-i \int^{1}_0\frac{d\tau }{\Lambda ^{2}}exp\{\frac{\Im \tau}{ \Lambda ^{2}}\}$$
In the context of the $\lambda \phi^4$ theory, we have,
$$\Im = -\partial^2-m^2 \quad I(\phi)= -\frac{\lambda}{4}  \phi^4$$
We shall now make elaborative comments.
The procedure constructs an action having an infinite number of terms (each individually local), and  having  arbitrary order derivatives of $\phi$.  The net result is to give convergence in the \emph{Euclidean} momentum space beyond a momentum scale $\Lambda$ through a factor of the form $\exp{(\frac{p^2-m^2}{\Lambda ^2})}$ in propagators. The construction is such that there is a one-to-one correspondence between the solutions of the local and the non-local classical field equations, (a difficult task indeed \cite{EW}). It can also be made to preserve the local symmetries of the local action in a non-localized form \cite{KW92}.

The Feynman rules for the nonlocal theory are simple extensions of the local
ones. In momentum space, these read:
\begin{enumerate}

\item For the $\phi $-propagator (smeared propagator) denoted by a straight line: $i\frac{\left\{ exp\left[\frac{p^{2}-m^{2}+i\epsilon}{\Lambda ^{2}}\right]\right\} }{p^{2}-m^{2}+i\epsilon}=\frac{-i}{\Lambda^2}\int_{1}^{\infty}d\tau\exp\left\{ \tau\left[\frac{p^{2}-m^{2}+i\epsilon}{\Lambda ^{2}}\right]\right\}$

\item For the $\psi $-propagator denoted by a \emph{barred} line:$$i\frac{\left\{ 1-exp\left[\frac{p^{2}-m^{2}+i\epsilon}{\Lambda ^{2}}\right]\right\} }{p^{2}-m^{2}+i\epsilon}=\frac{-i}{\Lambda^2}\int_{0}^{1}d\tau\exp\left\{ \tau\left[\frac{p^{2}-m^{2}+i\epsilon}{\Lambda ^{2}}\right]\right\}$$

\item The 4-point vertex is as in the local theory, except that any of the lines emerging from it can be of either type. (There is accordingly a statistical factor).

\item In a Feynman diagram, the internal
lines can be either shadow or smeared, with the exception that no diagrams can have closed shadow loops. \end{enumerate}
\begin{small}
We add several observations:
\begin{description}
    \item[a] The shadow propagator is the residue left behind when the smeared one is subtracted from the local one.
    \item[b] The smeared propagator has residue "1" at the location of pole, whereas the shadow propagator has no pole. These remarks are relevant to the unitarity of the theory, proven via Cutkosky rules.
    \item[c] The lack of pole in the shadow propagator is relevant in the discussion of mass-singularity in CV amplitude.
    \item [d] The parameter $\Lambda$ in the theory can be constrained from precision experiments [See e.g. first of \cite{js}and \cite{aj}] and theoretical arguments based on an understanding of renormalization in this framework \cite{j01}.
\end{description}
\end{small}

\subsection{Bogoliubov-Shirkov Condition of Causality}
The causality condition that we have used to investigate causality
violation in NLQFT is the one discussed by Bogoliubov and Shirkov \cite{BS}.
They have shown that an S-matrix for a theory that preserves causality
must satisfy the condition of Eq.(\ref{eq:CC}) and it has been formulated
treating the coupling $g(x)$ as space-time dependent.
A simple \emph{qualitative} understanding can be provided as follows:
Let us consider a theory with a variable coupling constant $g(x)$ and having time-reversal invariance. Then the S-operator satisfies:$S[g(z)]S^\dagger[g(z)]=\mathcal{I}$, independent of $g(z)$. $S^\dagger$ is an operator that evolves a state back in time from $t=\infty$ to $t=-\infty$, while $S$ then carries it back to $t=\infty$ giving the original state. In this to and fro time travel, any variation in $g(z)$, done on both routes does not change the net result, $\mathcal{I}$: the change in $g(z)$ on both routes together cancels out. Now suppose we carry out a local change in $g(z)$ only on the return trip at $y$: $\delta g(z)=0, z \neq y$. Then, $S[g(z)+\delta g(z)]S^\dagger[g(z)]\neq \mathcal{I}$ $= S[g(z)]S^\dagger[g(z)]$. Now, causality demands that the change in $g(z)$ at $y$ \emph{cannot affect the time evolution for times earlier than $y$}. Thus, the time-evolution on the return path between $(-\infty,y_0)$ is exactly reverse of that on the first part between $(y_0,-\infty)$ and thus  cancels out in the combination $S[g(z)+\delta g(z)]S^\dagger[g(z)]$. Hence, in the combination $S[g(z)+\delta g(z)]S^\dagger[g(z)]$, any additional change in $g(z)$ in the vicinity of any point $x$ \emph{earlier} than $y$ and \emph{made on both routes}, would cancel out similarly between the two trips. Hence,
$$\frac{\delta }{\delta g(x)}\left[S[g(z)+\delta g(z)]S^\dagger[g(z)]\right]=0$$
for $x_0 < y_0$. Also, if $x\sim y$, i.e. $x$ spacelike with respect to $y$, then there exists a frame in which $x_0 < y_0$ holds. Covariance then leads to the same relation in this case also.\\
The above relation is a series in $g(x)$ and leads perturbatively to an infinite set of equations when expanded using
\begin{equation}
S[g]=1+\sum _{n\geq 1}\frac{1}{n!}\int S_{n}(x_{1},...,x_{n})g(x_{1})...g(x_{n})dx_{1}...dx_{n}.\label{eq:CCexp}\end{equation}
These can be further simplified by the use of unitarity relation $S^\dagger[g(x)] S[g(x)]=\mathcal{I}$, expanded similarly in powers of $g(x)$.\\
We shall  write only a few of each of these (for a general expression for $H_n$, see \cite{JJ}):
\begin{equation}
H_{1}(x,y)\equiv iS_{2}(x,y)+iS_{1}(x)S_{1}^{\dagger }(y)=0\label{causal1}\end{equation}
\begin{equation}
H_{2}(x,y,z)\equiv iS_{3}(x,y,z)+iS_{1}(x)S_{2}^{\dagger }(y,z)+iS_{2}(x,y)S_{1}^{\dagger }(z)+iS_{2}(x,z)S_{1}^{\dagger }(y)=0\label{causal2}\end{equation}
(valid for $x_0>y_0,z_0$); along with the unitarity condition gives by\begin{equation}
S_{1}(x)+S_{1}^{\dagger }(x)=0\label{unitary1}\end{equation}
 \begin{equation}
S_{2}(x,y)+S_{2}^{\dagger }(x,y)+S_{1}(x)S_{1}^{\dagger }(y)+S_{1}(y)S_{1}^{\dagger }(x)=0\label{unitary2}\end{equation}
In the
case of the local theory, these relations [(\ref{causal1}) and (\ref{causal2})] are trivially satisfied.
In the case of the nonlocal theories, such quantities, on the other
hand, afford a way of characterizing the causality violation. However,
these quantities contain not the usual S-matrix elements that one
can observe in an experiment (which are obtained with a \emph{constant
i.e. space-time-independent} coupling), but rather the coefficients
in (\ref{eq:CCexp}). We thus find it profitable to construct appropriate
space-tine integrated versions out of $H_{n}(y,x_{1},...,x_{n})$.
Thus, for example, we can consider
\begin{eqnarray}
H_{1}&\equiv &\int d^{4}x\int d^{4}y[\vartheta (x_{0}-y_{0})H_{1}(x,y)+\vartheta (y_{0}-x_{0})H_{1}(y,x)]\nonumber \\
&=& i\int d^{4}x\int d^{4}y S_{2}(x,y)-i\int d^{4}x\int d^{4}y T[S_{1}(x)S_{1}(y)]\label{eq:H_1}\end{eqnarray}
 which can be expressed entirely in terms of Feynman diagrams that
appear in the usual S-matrix amplitudes. In a similar manner, we can
formulate
\begin{eqnarray}
H_{2} & \equiv &  \int d^{4}x\int d^{4}y\int d^{4}zH_{2}(x,y,z)\vartheta (x_{0}-y_{0})\vartheta (y_{0}-z_{0}) \\
&  &      +5\,\mbox{symmetric}\; \mbox{terms}\label{eq:H_2}
\end{eqnarray}
and can itself be expressed in terms of Feynman diagrams.

\subsection{Classical and Quantum violations of Causality}
In this section, we shall discuss the essential dissimilarities  between the violation of causality at the classical and quantum levels\footnote{The discussion below is qualitative and broad enough but may not cover all possibilities.}. Suppose the theory has classical violation of causality with coupling strength $g^2$ and range $R\equiv\frac{1}{\Lambda}$. We may, for example, consider an action-at-a-distance signal the travels  \footnote{Such a possibility is inherent in a covariant non-local interaction with a mass parameter $\Lambda$.} a distance $\leq R$ and strength $g^2$. We may then consider a succession of a set of $N$ stationary transmitters placed a distance $R$  apart. This arrangement will transmit an action-at-a-distance signal of strength $\sim g^{2N}$. Assuming that the smallest observable amplitude is $\alpha$, the system transmits a causality-violating signal to a range $\sim \frac{\ln\alpha}{2\ln g}R$ which can be large even for a fixed $R$ and is being transmitted even at zero momentum of transmitters. On the other hand, for a quantum violation, there is first a further suppression of $\frac{g^2}{16\pi^2}$ per loop, thus cutting down the amplitude and the range. We shall, in addition, show that (as would be expected for a quantum phenomenon) the degree of quantum violation depends sensitively on the de Broglie wavelength $\lambda=h/p$ of the interacting particles and the causality violating effect is greatly diminished for $R<<\lambda$ and becomes significant only when $\lambda\sim R=1/\Lambda$, i.e. at high energies. In this connection, we note that in the classical limit $h\rightarrow 0$, the de Broglie wavelength $\lambda\rightarrow 0$ and always fulfills the condition for a large causality violation: $R\geq\lambda$ for any small momentum.
In absence of an evidence of a large-scale violation of causality, it is desirable to restrict to theories with only quantum violations of causality.

\section{Calculation of CV at one loop\label{sec:3}}

\vspace{0.3cm}
\begin{center}\includegraphics[width=4in,
  height=3in] {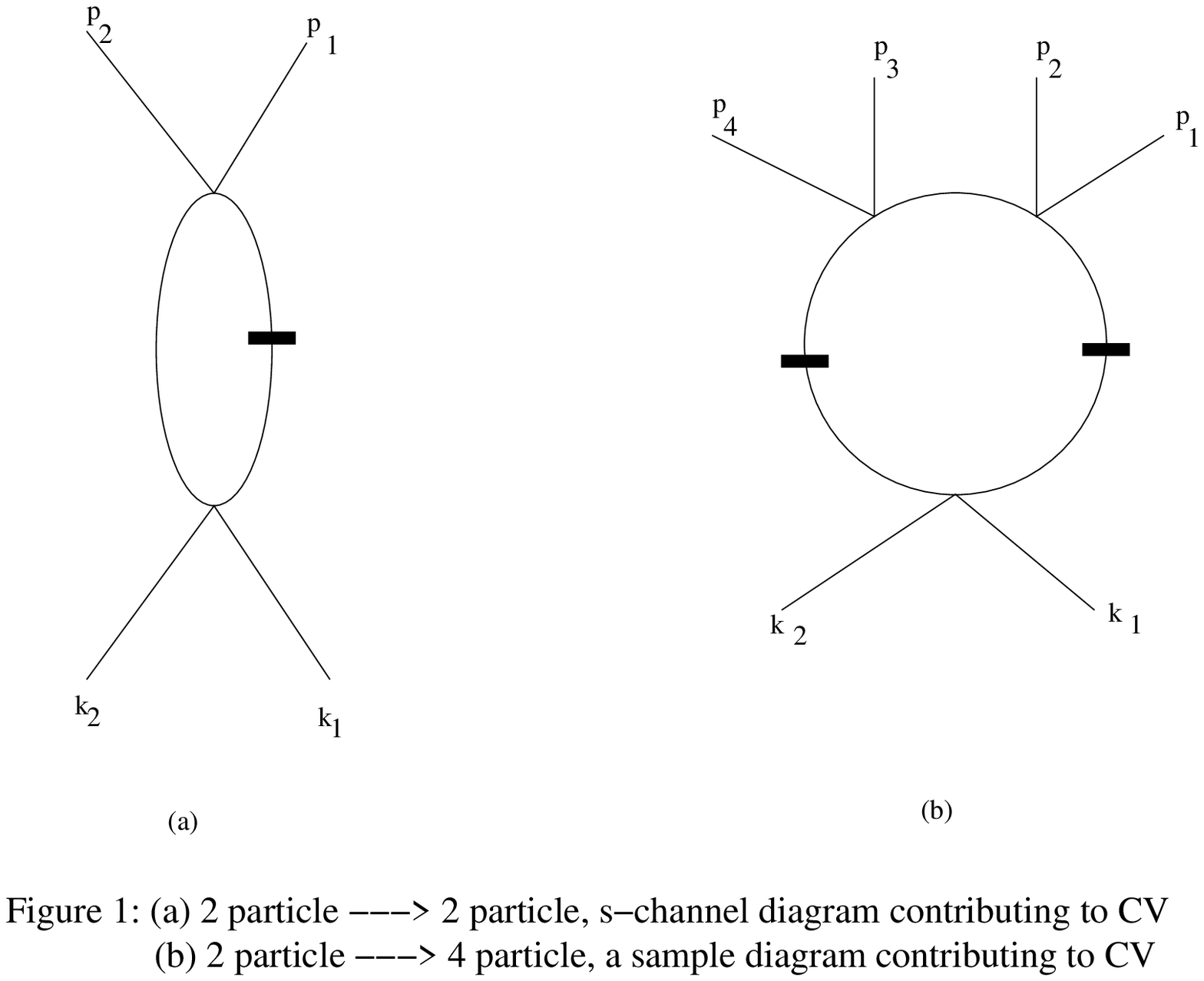}\end{center}
\vspace{0.3cm}

\subsection{Process $ \phi+\phi \rightarrow \phi+\phi$}
General properties of the amplitude for this process have been studied to all orders \cite{JJ} and the conclusions are presented at the end of this subsection. We shall first go into the details of the one-loop contribution. At one-loop, this process is of $O(g^2)$ and $H_1$ is the operator relevant for CV amplitude. The contribution to $H_1$ comes from two sources:
\begin{small}\begin{itemize}
    \item The diagram of Fig.1(a) and crossed diagrams
    \item The counter-terms inherent in each of $\int d^{4}x\int d^{4}y S_{2}(x,y)$  \\and  $\int d^{4}x\int d^{4}y T[S_{1}(x)S_{1}(y)]$
    \end{itemize}\end{small}
We give the result for the amplitude for the s-channel
diagram (Fig.1(a)) in the massless limit:
\begin{equation}
\Gamma _{0}(s)= \frac{9\lambda ^2}{4\pi^2} \sum _{n=0}^{\infty }\frac{\left(\frac{s}{\Lambda ^{2}}\right)^{n}\left(1-\frac{1}{2^{n}}\right)}{n((n+1)!)}.\label{eq:schannel}  \end{equation}
In addition, there are the $t$-channel and the $u$-channel diagrams each of which is given by an analogous expression with $s \rightarrow t $ and $s \rightarrow u $ respectively. The sum of three such diagrams, for the case $s \ll \Lambda ^2 $ is given by,
\begin{eqnarray}
\Gamma _{0}(s)+\Gamma _{0}(t)+\Gamma _{0}(u)& = &  \frac{9\lambda ^2}{4\pi^2}\left\{3 ln 2+ \frac{1}{4\Lambda^2}(s+t+u) +O\left(   \frac{s^2}{\Lambda^4 } \right)\right\}\nonumber  \\
& = & \frac{9\lambda ^2}{4\pi^2}\left\{3 ln 2+\frac{1}{\Lambda^2}(m^2) +O\left( \frac{s^2}{\Lambda^4 },\frac{t^2}{\Lambda^4 },\frac{u^2}{\Lambda^4 } \right)\right\}
\end{eqnarray}
The counter-terms, on the other hand, are partly determined by the requirement that the net CV $\rightarrow 0$ as $\Lambda\rightarrow\infty$. This still leaves an ambiguity of a constant that vanishes as $\Lambda\rightarrow\infty$. Consistent with phenomenological \emph{non-observation} of CV, we shall assume that CV vanishes at low energies (as $s\rightarrow 0$) \cite{JJ}, which will fix the ambiguity\footnote{As emphasized in \cite{JJ}, presence of any \emph{momentum-dependent} contribution to $H_1$ signals causality violation, irrespective of this assumption.}. The net result for CV then is,
\begin{equation}
\frac{9\lambda ^{2}}{64{\pi}^2}\left(\frac{s^{2}+t^{2}+u^{2}}{\Lambda ^{4}}\right).\end{equation}
Thus, the CV is an order of magnitude smaller than would be expected by naive power counting.
On the other hand, for large enough $s \leq \Lambda^2$, the series for $\Gamma_0(s)$ shows a rapid (exponential-like) increase, while the series for $\Gamma_0(t)$ and $\Gamma_0(u)$ die away. The CV then shows a rapid exponential-like rise.

These results for ($ \phi+\phi \rightarrow \phi+\phi$), arrived at from a relatively simple 1-loop calculation \emph{can} be generalized to \emph{all orders} \cite{JJ}.  They are enumerated below:
\begin{small}\begin{itemize}
    \item $CV\propto \left(\frac{s^{2}+t^{2}+u^{2}}{\Lambda ^{4}}\right)$ for $s<<\Lambda^2$
    \item CV shows a rapid (exponential-like)  rise as $s \leq \Lambda^2$
    \item CV shows no mass singularity as $m\rightarrow0$. This is important in a theory having small masses (e.g.$m_e,m_{\nu}$). This result forbids a large CV developing $\sim ln(m)$.
    \item CV amplitude is real: the graphs contributing to it have no physical intermediate states.
    \end{itemize}    \end{small}
The proof depends on the analyticity properties of the CV amplitudes in the variables $s,t$ and $m$. Very compact arguments can be constructed to study these\footnote{See e.g. the Appendix A of \cite{JJ}.}.
\subsection{Process $ \phi+\phi \rightarrow \phi+\phi+\phi+\phi$}
A similar calculation can be made for $H_2$ in the one loop approximation. The diagram of Fig. 1(b), together with 14 others obtained by Bose-symmetry, contribute. There is no counterterm here. The results are identical:
\begin{small}\begin{itemize}
    \item $\frac{CV}{\mbox{total amplitude}} \sim \Lambda^{-2}\times\{\mbox{a quadratic function of Lorentz invariants}\}$\\ as would be expected by power counting.
    \item CV shows a rapid (exponential-like) rise for $s \leq \Lambda^2$
\end{itemize}      \end{small}
For further details, please see \cite{JJ}.

\section{Interpretation of Results and further directions}
The above calculations predict causality violation  in the  NLQFT model, especially a larger CV amplitude for higher energies $s\sim \Lambda^2$. We recall that experiments (g-2 of the muon, precision tests of the standard model) constrain $\Lambda$ [See e.g. first of \cite{js},\cite{j01} and \cite{aj}]. As elaborated in section 1, the NLQFT's have (at least) two possible interpretations:\begin{small}\begin{itemize}
    \item [I.] $1/\Lambda$ represents scale of non-locality that determines "granularity" of space-time. Then $1/\Lambda$ is a fixed property of space-time for any field theory.
    \item [II.] The non-local theory represents an effective field theory and the scale $\Lambda$  represents the scale at which the theory has to be replaced by a more fundamental theory.
\end{itemize} \end{small}
We can understand/interpret these results in both the frameworks, but the meaning attached to them is different.
Option I necessarily requires a relatively large causality violation at  $s \sim \Lambda^2$ . An observation of causality violation at these energies will bolster an interpretation of these theories as a physical theory with first interpretation.
In this picture, for low energies, the De Broglie wavelength $ \lambda$ is much larger than the space-time scale of non-locality, and causality violation would go unobserved. On the other hand, for energies $s\sim\Lambda^2$,
$\lambda\sim h/\Lambda$, the scale of non-locality. So it is not surprising if CV becomes significant.
Option II leaves the possibility that as $s \rightarrow\Lambda^2$, the non-local theory becomes less and less valid; because then we should have to use the underlying (fundamental) theory to calculate quantities. In this case, the large CV obtained by calculation \emph{could} be an artifact of approximation that replaces the more fundamental theory by an effective non-local theory. Thus, a non-observation of a growing CV amplitude, could possibly be re-interpreted by going to the underlying fundamental theory and is not necessarily in an immediate contradiction with the above calculations.\\
Work in several directions to generalize the results is in progress.

\begin{small}

\end{small}


\begin{thebibliography}{10}
\bibitem {PU} A. Pais and G. E. Uhlenbeck, Phys. Rev. 79, 145-165 (1950).
\bibitem{NA} See, e.g. K. Namsrai, \emph{Nonlocal Quantum Field Theory and Stochastic
Quantum Mechanics} (D. Reidel Publishing company).
\bibitem{M90} J.Moffat Phys. Rev. D41,1177(1990)
\bibitem{NC} See e.g. N. Seiberg, Leonard Susskind, N. Toumbas; JHEP 0006:044,(2000) and references therein.
\bibitem{KW92} G. Kleppe and R. P. Woodard, Nucl. Phys. B388, 81 (1992)
\bibitem{EW} D. A. Eliezer, and R. P. Woodard, Nucl.Phys.B 325, 389 (1989).
\bibitem{E91} E. D. Evens et al, Phys Rev D43, 499 (1991)
\bibitem{j01} S.D.Joglekar, J. Phys. A34, 2765-2776 (2001)
\bibitem{j01_2} S.D.Joglekar,Int.J.Mod. Phys.A 16, (2001).
\bibitem{js} S. D. Joglekar and G. Saini, Z. Phys. C.76, 343-353 (1997);  A. Basu, and S. D. Joglekar, J. Math. Phys. 41, 7206-7219 (2000); A.
Basu, and S. D. Joglekar, Euro.Phys. Journal-direct C4 3, 1-20 (2001).
\bibitem{CO92}N. J. Cornish, Int.J.Mod. Phys.A 7, 6121-6157 (1992).
\bibitem{JJ} A. Jain and S.D.Joglekar, Int.J. Mod. Phys.A 19, 3409(2004)
\bibitem{BS}N.N.Bogolibov and D.V.Shirkov, `Introduction to Theory of Quantized Fields' ($3^{rd}$ ed.), John Wiley(1980). See pg. 200-220.
\bibitem{aj} A.Ayyer and S.D.Joglekar [unpublished calculations]
\end{thebibliography}
\end{document}